\documentclass[submission,Proceedings]{SciPost}
\newcommand{\emu}{{\left(\frac{\partial{\epsilon}}{\partial{\mu}}\right)_{T, B}}}
\newcommand{\eT}{{\left(\frac{\partial{\epsilon}}{\partial{T}}\right)_{\mu, B}}}

\usepackage[normalem]{ulem}

\usepackage{amsmath}
\usepackage{xcolor}

\newcommand{\rhomu}{{\left(\frac{\partial{\rho}}{\partial{\mu}}\right)_{T, B}}}
\newcommand{\rhoT}{{\left(\frac{\partial{\rho}}{\partial{T}}\right)_{\mu, B}}}

\begin{document}

\begin{center}{\huge \textbf{How to sit Maxwell and Higgs\\ on the boundary of Anti-de Sitter
}}\end{center}

\begin{center}
\textbf{Matteo Baggioli} \textsuperscript{1,2*}
\end{center}

\begin{center}
{\bf 1} School of Physics and Astronomy, Shanghai Jiao Tong University, Shanghai 200240, China
\\
{\bf 2} Wilczek Quantum Center, School of Physics and Astronomy, Shanghai Jiao Tong University, Shanghai 200240, China

\end{center}

\begin{center}
*  \href{mailto:b.matteo@sjtu.edu.cn}{b.matteo@sjtu.edu.cn} 
\end{center}

\begin{center}
\today
\end{center}


\section*{Abstract}
{\bf
In the context of bottom-up holography, we demonstrate the power of mixed boundary conditions to promote the boundary gauge field to be dynamical. We provide two concrete applications of this idea. First, we consider a holographic dual for a strongly coupled plasma described by dissipative magnetohydrodynamics. Second, we reveal the expected features of the Higgs mechanism in a not counterfeit holographic superconductor.
}

\vspace{0.5cm}

\noindent \textit{Proceedings for the 6$^{th}$ International Conference on Holography, String Theory and Spacetime, held in Da Nang (Vietnam), 20-24 February 2023 (\href{https://duytan.edu.vn/conferences/home/CallForPaper/12-6th-international-conference-on-holography-string-theory-and-spacetime-in-da-nang}{website})\\\\
\noindent Talk based on \cite{Ahn:2022azl,Jeong:2023las} in collaboration with Yongjun Ahn, Kyoung-Bum Huh, Hyun-Sik Jeong, Keun-Young Kim and Ya-Wen Sun.}


\section{Introduction}
In every introductory lecture on the AdS-CFT correspondence, a commonly repeated slogan is that ``gauge symmetries in the gravitational bulk correspond to global symmetries in the boundary dual field theory" \cite{Maeda:2010br}. Such a folkloristic statement immediately brings the question of whether one would ever be able to holographically describe systems in which gauge degrees of freedom, and the associated gauge interactions, are not only unavoidable but fundamental. The simplest example of this sort is certainly the case of electromagnetism, i.e., a U(1) gauge symmetry. In other words, can we have a holographic model whose dual field theory contains a dynamical gauge field, a propagating photon, and electromagnetic interactions? And if yes, how? These questions are not merely academic. Indeed, in view of ``practical'' and realistic applications of holography, especially in the context of condensed matter, they represent important obstacles. In a way, this situation is reminiscent of the similar need to introduce momentum relaxation and translational symmetry breaking in holographic bottom-up models, which has sparkled a huge amount of effort and produced important results in the last decade \cite{Baggioli:2022pyb}.

Fortunately, the answers to the questions above were promptly given long time ago by Witten \cite{Witten:1998qj,w2}, and later formalized for the specific case of bulk gauge fields by Marolf and Ross \cite{Marolf:2006nd}. Let us consider a $(d+1)$-dimensional spacetime metric which is asymptotically Anti de Sitter. To set our notations, let us use a holographic radial coordinate such that the boundary is located at $r=\infty$. Then, the asymptotically UV behavior of a generic bulk field $A_\mu(t,\vec{x},r)$ is given by:
\begin{equation}\label{exp}
A_\mu\left(t,\vec{x},r\right)\,=\,A^{(0)}_\mu\left(t,\vec{x}\right)\,+\,A^{(1)}_\mu\left(t,\vec{x}\right)\,r^{2-d}+\dots 
\end{equation}
The most general boundary condition for the bulk gauge field corresponds to fixing the linear combination
\begin{equation}\label{bcs1}
\alpha\,A^{(0)}_\mu\left(t,\vec{x}\right)\,+\,\beta\,A^{(1)}_\mu\left(t,\vec{x}\right) \,,
\end{equation}
and takes the name of \textit{mixed boundary condition} (or Robin boundary condition).

\begin{figure}
    \centering
    \includegraphics[width=0.85\linewidth]{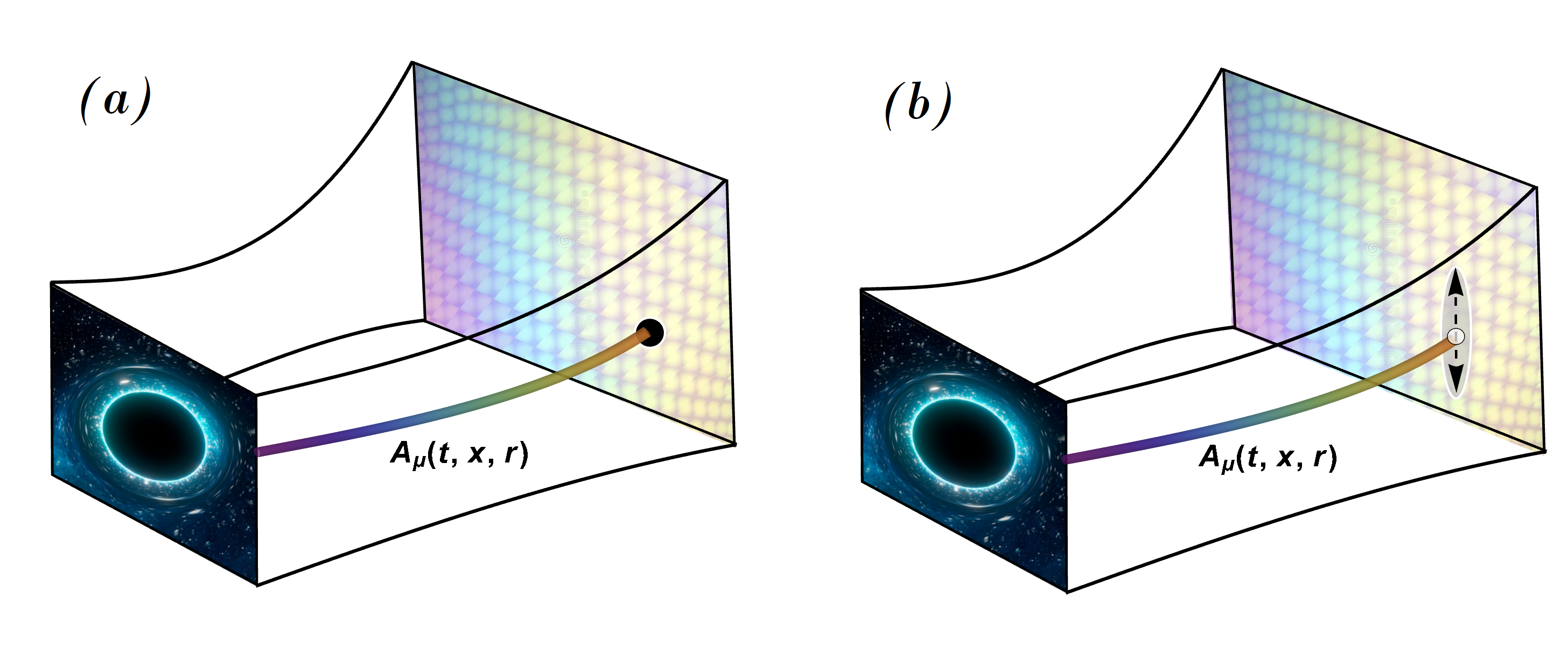}
    \caption{A cartoon for the different boundary conditions. \textbf{(a)} The standard Dirichlet boundary condition for the bulk gauge field $A_\mu(t,x,r)$. \textbf{(b)} The mixed boundary condition for the bulk gauge field $A_\mu(t,x,r)$.}
    \label{fig0}
\end{figure}

The standard Dirichlet case (see cartoon in Fig.\ref{fig0}), $\beta=0$, is equivalent to deform the boundary CFT as:
\begin{equation}
\mathcal{L}_{CFT}\,\,\longrightarrow\,\,\mathcal{L}_{CFT}\,+\,\int d^dx\,A^{(0)}_\mu J^\mu \,,
\end{equation}
where $J^\mu$ is a U(1) current operator of the dual CFT with conformal dimension $\Delta=d-1$ and $A^{(0)}_\mu$ a fixed external gauge field which acts as a source for it. As a result of this quantization choice, the dual field theory does not have a dynamical gauge field in its spectrum, and any electric or magnetic fields considered are external. Within this quantization scheme, the holographic setup provides us with the generating functional of the dual field theory as a function of the external sources, $g_{\mu\nu}$ and $A_\mu$,
\begin{equation}\label{EXTTH}
  Z[g_{\mu\nu},A_\mu]= \int \mathcal{D}\Phi \exp \left[i S_0\left(\Phi\right)+ i \int d^d x \sqrt{-g} \left(A_\mu J^\mu\left(\Phi\right)+\frac{1}{2}g_{\mu\nu}T^{\mu\nu}\left(\Phi\right)\right)\right] \,,
\end{equation}
where $\Phi$ is just a collective label for all the dynamical fields of the dual CFT. From the generating functional in Eq.\eqref{EXTTH}, we can define the effective action as
\begin{equation}\label{EXTTH2}
    S[g_{\mu\nu}, A_\mu] := -i \, \ln Z[g_{\mu\nu}, A_\mu]\,,
\end{equation}
and extract all the correlation functions for the operators $T^{\mu\nu}$ and $J^\mu$ coupled to our external sources, e.g., $\langle J^\mu J^\nu \rangle$, $\langle T^{\mu\nu}J^\sigma J^\delta\rangle$, etc.
As discussed at length in \cite{Maeda:2010br,Montull:2009fe,Domenech:2010nf}, imposing a more general boundary condition as in Eq.(\ref{exp}) allows for the possibility to have an emergent dynamical gauge field in the dual field theory. Formally, this procedure corresponds to modify the effective action as follows
\begin{equation}
\label{DYNATH}
    S_{\text{tot}}\,= S_{\text{m}}[g_{\mu\nu}, A_\mu] \,+\, \int d^d x\sqrt{-g} \,\left[\,\color{red}-\frac{1}{4\lambda}F^2\, + A_\mu \, J^{\mu}_{\text{ext}}\color{black}\right]\,.
\end{equation}
The new pieces, shown in red color in the equation above, correspond to the kinetic term for the boundary gauge field and the Legendre transform in terms of the external current $J^{\mu}_{\text{ext}}$. Here, $\lambda$ is the square of the electromagnetic coupling. By varying the total action \eqref{DYNATH} with respect to the dynamical field $A_\mu$, one obtains
\begin{equation}\label{SUMBDY333}
\delta_{A_\mu} S_{\text{tot}} \,=\, \int d^3x \, \sqrt{-g} \,\, \left[ J^{\mu}_{\text{m}} \,-\, \frac{1}{\lambda} \nabla_{\nu} F^{\mu\nu} + J_{\text{ext}}^{\mu}  \right]  \delta A_{\mu}   \,,
\end{equation}
which gives the expected Maxwell equations.

Remaining in the field theory side, the complete set of equations of motion are given by
\begin{align}\label{}
&\nabla_\mu \left(T^{\mu \nu}_{\text{m}} \,+\, T^{\mu \nu}_{\text{EM}}\right) \,\,=\, F^{\lambda \nu}J_{\text{ext} \lambda} \,, \qquad\qquad\,\,\,  \nabla_\mu J^\mu_{\text{m}} \,=\, 0 \,,  \label{Eq:MHDdyn1} \\
&J^{\mu}_{\text{m}} \,+\, J^{\mu}_{\text{EM}} \,+\, J_{\text{ext}}^{\mu} = 0\,, \qquad \,\,\,\,   \qquad \quad \epsilon^{\alpha \beta \gamma} \, \nabla_\alpha F_{\beta \gamma} \,=\, 0 \,. \label{Eq:MHDdyn2}
\end{align}
In there, we have separated the matter and EM contributions to the stress-energy tensor and current which are defined as follows:
\begin{align} \label{CONJ3}
&\delta_{g_{\mu\nu}} S_{\text{m}} \,=\, \frac{1}{2} \int d^3 x \,\, \sqrt{-g} \,\,  T^{\mu\nu}_{\text{m}} \, \delta g_{\mu\nu} \,,\quad  \delta_{A_\mu} S_{\text{m}} \,=\, \int d^3 x \,\, \sqrt{-g} \,\, J^{\mu}_{\text{m}} \, \delta A_{\mu} \,\\
& T^{\mu\nu}_{\text{EM}}=\frac{1}{\lambda}F^{\mu\sigma}F^{\nu}_{\,\,\,\,\sigma}-\frac{1}{4\lambda}F^2 g^{\mu \nu}\qquad J^{\mu}_{\text{EM}}=-\frac{1}{\lambda}\nabla_\nu F^{\mu\nu}\,.
\end{align}
Using the equations above, supplemented with the correct constitutive relations written in the proper gradient expansion, one can formally derive a low-energy effective description known as relativistic magnetohydrodynamics 
\cite{Hernandez:2017mch}. In such a framework, the dispersion relations of the low-energy modes can be consistently obtained as a function of the various transport coefficients.

Despite the importance of mixed boundary conditions to promote the boundary gauge field (and eventually also the boundary metric) to be dynamical has been early recognized and used in several situations, the study of strongly coupled plasmas and the related magneto-hydrodynamic phenomenology has never been performed so far, at least in the language of bulk gauge fields (see \cite{Grozdanov:2017kyl} for a higher-forms realization). In the following, we show how such an analysis can be performed using the b.c.s. in Eq.\eqref{bcs1}, in perfect agreement with the magneto-hydrodynamic theory \cite{Hernandez:2017mch}, and without the need of introducing higher-form bulk fields (see \cite{DeWolfe:2020uzb} for a nice explanation about the equivalence of the two approaches). Afterwards, we present how the same modified b.c.s. can help us to promote the holographic superfluid model \cite{Hartnoll:2008vx} into a \textit{bona-fide} holographic superconductor model, whose properties are in good agreement with Ginzburg-Landau theory and previous perturbative computations \cite{Natsuume:2022kic}.

\section{Magnetohydrodynamics from holography}
\subsection{Setup}
We consider the Einstein-Maxwell action in four spacetime dimensions
\begin{equation}\label{ACTIONH}
S_{\text{bulk}} = \int d^4x\, \sqrt{-g} \,\left( R \,+\, 6 \,-\, \frac{1}{4} F^2 \right) \,, \qquad F= d A \,,
\end{equation}
together with the dyonic black-brane ansatz 
\begin{equation}\label{BGMET}
d s^2 =  -f(r)\, d t^2 +  \frac{1}{f(r)} \, d r^2  + r^2 (d x^2 + d y^2) \,,\quad   A = A_t(r) \, d t -\frac{B}{2} y \,d x \,+\, \frac{B}{2} x \, d y \,,
\end{equation}
where $B$ is the magnetic field. A simple Reissner-Nordstrom solution can be obtained as
\begin{equation}\label{BCF}
\begin{split}
 f(r) &\,= r^2 - \frac{m_{0}}{r} \,+ \, \frac{\mu^2 r_{h}^2 + B^2}{4\,r^2} \,, \quad m_{0} = r_{h}^3\left( 1 +  \frac{\mu^2 r_{h}^2+B^2}{4\, r_{h}^4} \right) \,, \\
 A_{t}(r) &\,= \mu \left( 1- \frac{r_{h}}{r} \right)\,,
\end{split}
\end{equation}
where $\mu$ is the chemical potential, $r_{h}$ the horizon radius, and the mass $m_{0}$ is determined by the condition $f(r_{h})=0$.

We identify the bulk on-shell action in Eq.\eqref{ACTIONH} with the matter contribution $S_{m}[g_{\mu\nu}, A_{\mu}]$ in Eq.\eqref{DYNATH}. Moreover, we add the following boundary terms
\begin{equation}\label{Action:bdry}
S_{\text{boundary}} = \int d^3x \, \left[ -\frac{1}{4\lambda}F_{\mu\nu}^2 \,+\,  A_{\mu} \, J_{\text{ext}}^{\mu}  \right] \,,
\end{equation}
exactly as done in the field theory side in Eq.\eqref{DYNATH}.
The various thermodynamic quantities in the dual field theory are then given by
\begin{align}\label{HAWKINGT}
\begin{split}
 T &\,=\, \frac{1}{4\pi} \left( 3\,r_{h} \,-\, \frac{\mu^2 r_{h}^2 + B^2}{4\,r_{h}^3}  \right) \,,  \quad \rho \,=\, \mu \, r_{h} \,, \quad s \,=\, 4\pi \, r_{h}^2  \,, \\
\epsilon &\,=\,  2r_{h}^3 + \frac{\mu^2 r_{h}}{2} + \frac{B^2}{2 r_{h}} {+\frac{B^2}{2\lambda}}  \,, \qquad p \,=\, r_{h}^3 + \frac{\mu^2 r_{h}}{4} - \frac{3 B^2}{4 r_{h}} {-\frac{B^2}{2\lambda}}\,,   
\end{split}
\end{align}
where $(T,\rho,s, \epsilon, p)$ are the temperature, charge density, entropy, energy and pressure density, respectively.

On top of our background, we consider the following fluctuations
\begin{align}\label{}
\begin{split}
g_{MN} \,\rightarrow\, g_{MN} + \delta g_{MN} \,, \quad A_{M} \,\rightarrow\, A_{M} + \delta A_{M} \,,
\end{split}
\end{align} 
and for convenience we assume the radial gauge, in which all components with at least one ``$r$'' index are set to vanish. After Fourier transforming the various fields, we define the following gauge-invariant variables
\begin{align}\label{GIVOUR}
\begin{split}
Z_{H_1} &\,:=\, k \,h_{t}^{y} \,+\, \omega \, h_{x}^{y}  \,, \\
Z_{H_2} &\,:=\, \frac{4 k}{\omega} \, h_{t}^{x} \,+\,  2 h_{x}^{x} - \left( 2 - \frac{k^2}{\omega^2}\frac{f'(r)}{r} \right) h_{y}^{y}  + \frac{2k^2}{\omega^2}\frac{f(r)}{r^2} h_{t}^{t}  \,, \\
Z_{A_1} &\,:=\, k \,a_{t} \,+\, \omega \, a_{x} \,-\, \frac{i B\,\omega}{k} h_{x}^{y} \,-\,\frac{k\,r}{2} \,A_{t}' \, h_{y}^{y} \,, \\
Z_{A_2} &\,:=\, a_{y}  \,+\, \frac{i B}{2k} \left(h_{x}^{x}-h_{y}^{y}\right) \,,
\end{split}
\end{align}
where the index of the metric fluctuation $h_{MN}$ is raised with the background metric \eqref{BGMET}. Note that, near the AdS boundary, the gauge-invariant variables behave as
\begin{align}\label{BD2}
\begin{split}
&Z_{H_{i}} = Z_{H_{i}}^{(L)} \, r^{0} \,(1 \,+\, \dots) \,+\, Z_{H_{i}}^{(S)} \, r^{-3} \,(1 \,+\, \dots) \,, \\
&Z_{A_{i}} = Z_{A_{i}}^{(L)} \, r^{0} \,(1 \,+\, \dots) \,+\, Z_{A_{i}}^{(S)}\, r^{-1} \,(1 \,+\, \dots) \,,
\end{split}
\end{align}
where the superscripts $(L,S)$ denote the leading/subleading terms.

In order to determine the correct boundary conditions, we perform the variation of the total action $S_{\text{on-shell}}+S_{\text{boundary}}$, and obtain
\begin{equation} \label{SOLEQ2C}
\begin{split}
J^{\mu}_\text{m} \,-\, \frac{1}{\lambda} \partial_{\nu} F^{\mu\nu} + J_{\text{ext}}^{\mu} = 0 \,, \qquad J^{\mu}_\text{m} \,=\, \frac{\delta S_{\text{on-shell}}}{\delta A_{\mu}} \,=\, - \sqrt{-g}\, F^{r \mu} \big|_{r\rightarrow\infty} \,,
\end{split}
\end{equation}
where $S_{\text{on-shell}}$ is the action in \eqref{ACTIONH} computed on-shell. After algebraic manipulations, the above equation reduces to a set of conditions given by
\begin{equation}
\label{Eq:sourcess}
\begin{split}
&\delta J_{\text{ext}}^{\,t \,\,(L)} \,=  - \frac{k}{\lambda}Z_{A_1}^{(L)} - \frac{k}{\omega^2 - k^2}Z_{A_1}^{(S)} + \frac{\rho}{2(\omega^2-k^2)}Z_{H_2}^{(L)}\,, \\
&\delta J_{\text{ext}}^{\,x \,\,(L)} \,=  - \frac{\omega}{\lambda}Z_{A_1}^{(L)} - \frac{\omega}{\omega^2 - k^2}Z_{A_1}^{(S)} + \frac{\rho \omega}{2k(\omega^2-k^2)}Z_{H_2}^{(L)}\,, \\
&\delta J_{\text{ext}}^{\,y \,\,(L)} \,=  - \frac{\omega^2-k^2}{\lambda}Z_{A_2}^{(L)} - Z_{A_2}^{(S)}+\frac{\rho}{k}Z_{H_1}^{(L)}\,,
\end{split}
\end{equation}
where the l.h.s. refer to the sources in our boundary field theory, and $\omega,k$ are respectively the frequency and the wave-vector.

To compute the quasi-normal modes, we need to consider the determinant of the source matrix. This is given by
\begin{align}\label{APPENSMATA3}
\begin{split}
& \det \mathcal{S} = \left|
\begin{array}{cccc} 
Z_{H_1}^{(L)(I)} & Z_{H_1}^{(L)(II)} & Z_{H_1}^{(L)(III)} & Z_{H_1}^{(L)(IV)} \\ [8pt]
Z_{H_2}^{(L)(I)} & Z_{H_2}^{(L)(II)} & Z_{H_2}^{(L)(III)} & Z_{H_2}^{(L)(IV)} \\ [8pt]
\delta J_{\text{ext}}^{\,x \,\,(L)(I)} & \delta J_{\text{ext}}^{\,x \,\,(L)(II)} & \delta J_{\text{ext}}^{\,x \,\,(L)(III)} & \delta J_{\text{ext}}^{\,x \,\,(L)(IV)} \\ [8pt]
\delta J_{\text{ext}}^{\,y \,\,(L)(I)} & \delta J_{\text{ext}}^{\,y \,\,(L)(II)} & \delta J_{\text{ext}}^{\,y \,\,(L)(III)} & \delta J_{\text{ext}}^{\,y \,\,(L)(IV)}
\end{array}  \right| \,, \\
\end{split}
\end{align}
where $(I),(II),(III),(IV)$ indicate four linearly independent solutions of the equations of motion for the fluctuations. Because of the conservation equation $\nabla_\mu J_{\text{ext}}^{\mu}=0$, no additional condition on the time component $\delta J_{\text{ext}}^{\,t \,\,(L)}$ has to be imposed. Finally, the low-energy excitations in the dual field theory can be obtained by solving numerically:
\begin{equation}
    \det \mathcal{S}(\omega,k)=0
\end{equation}
as a function of $\mu,B,\lambda,T$ (or more precisely as a function of their dimensionless combinations).
\subsection{Showcase of (some of) the results}
In this section, we present some of the main results of our study. For more details, and an extensive review of all the results, we refer to \cite{Ahn:2022azl}.
\begin{figure}[]
\centering
     \includegraphics[width=6.3cm]{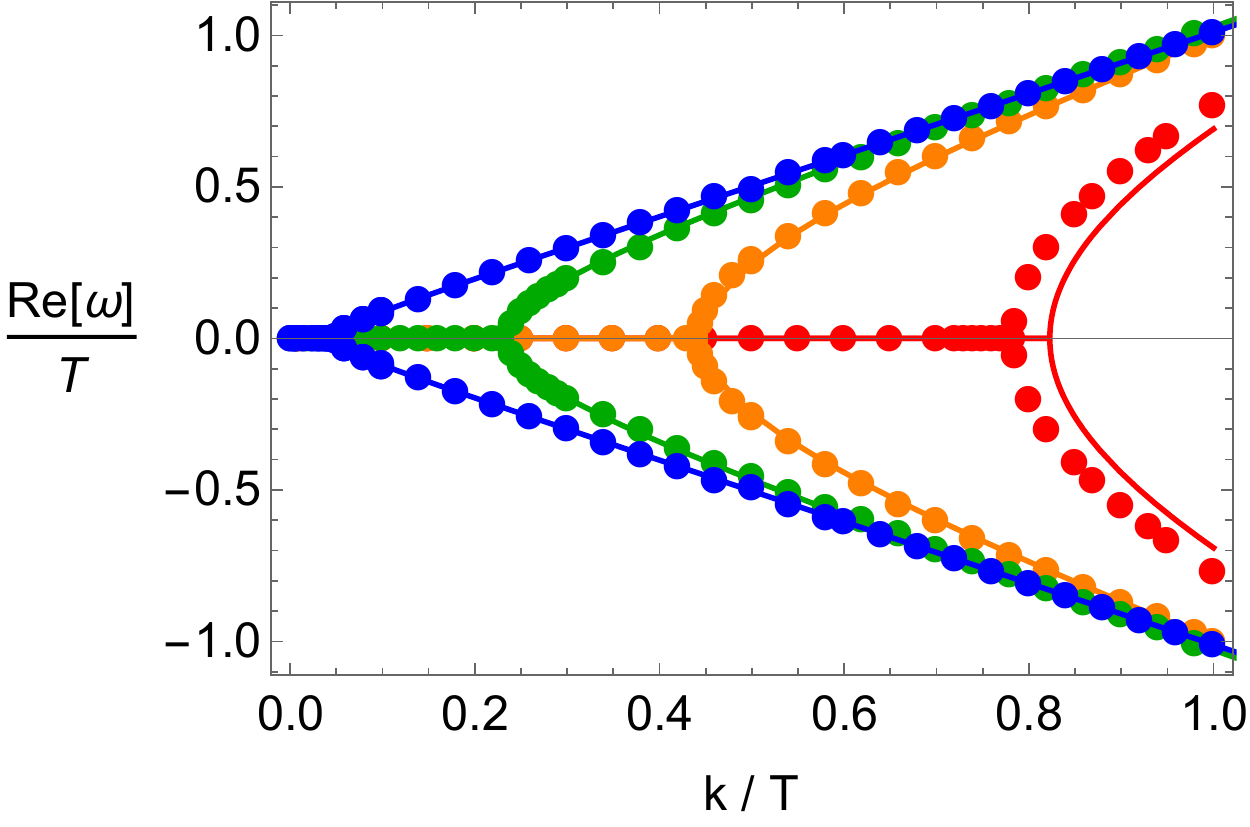} \qquad
     \includegraphics[width=6.3cm]{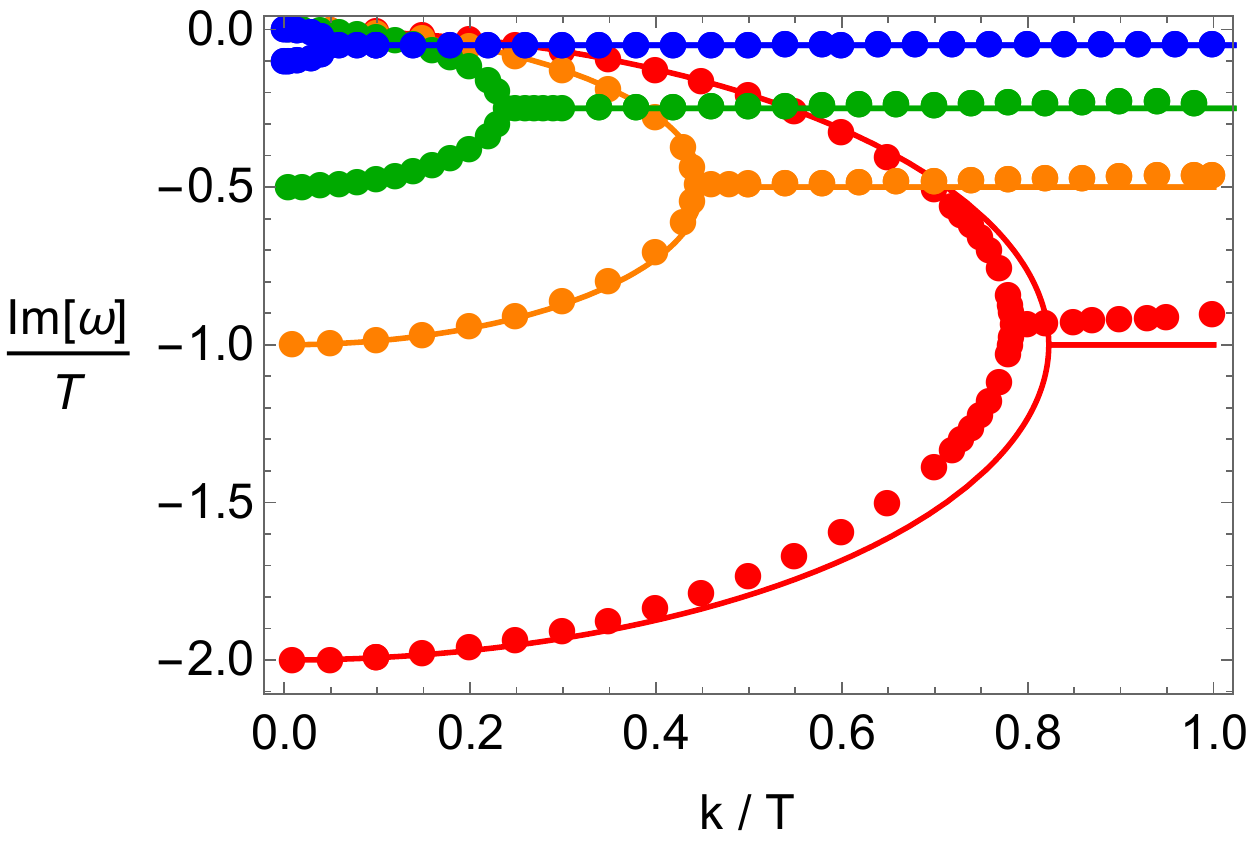} 
 \caption{The dynamics of the EM waves at zero density and zero magnetic field.  Different colors, from blue to red, correspond to $\lambda/T=0.1,0.5,1,2$. The symbols are the numerical results. The solid lines are the independent predictions from magnetohydrodynamics.}\label{fig1}
\end{figure}
As a first example, we consider the dynamics of the boundary field theory at zero chemical potential and magnetic field, $\mu=B=0$. In such a limit, the dynamics of the electric and magnetic fields decouples from the rest. In the EM sector, the corresponding low energy modes obey the following equation:
\begin{align}\label{EMWAVE}
\begin{split}
\omega \left(\omega + i \, \frac{\sigma}{\epsilon_{\text{e}}}\right) = \frac{k^2}{\epsilon_{\text{e}} \, \mu_{\text{m}}} \,,
\end{split}
\end{align} 
known as \textit{telegrapher} equation, and appearing in many areas of physics (see \cite{Baggioli:2019jcm} for a review). Here, $\sigma$ is the electric conductivity while $\epsilon_e,\mu_m$ are respectively the electric permittivity and the magnetic permeability. In vacuum, $\sigma=0$, and one recovers the standard EM wave (photon) with speed $c^2=1/(\epsilon_e \mu_m)$. In a dielectric material, or more in general whenever $\sigma \neq 0$, an imaginary term appears and destroys the presence of a propagating EM wave at low wave-vector. This is the reflection of the fact that electric lines are not anymore conserved and the electric field relaxes in time. On the contrary, magnetic lines are still conserved (or, in fancy words, the higher-form magnetic global symmetry is kept intact), giving rise to the magnetic diffusion mode
\begin{align}\label{EMWAVE2}
\omega = -i \frac{k^2}{\sigma \, \mu_{\text{m}}} +\dots\,
\end{align} 
The dynamics just described can be recovered within the holographic model as shown in Fig.\ref{fig1}. The numerical data are in perfect agreement with the magneto-hydrodynamic formulas. As expected, the screening effects, hindering the photon propagation, become larger by increasing the gauge coupling $\lambda$.
\begin{figure}[]
\centering

     {\includegraphics[width=7cm]{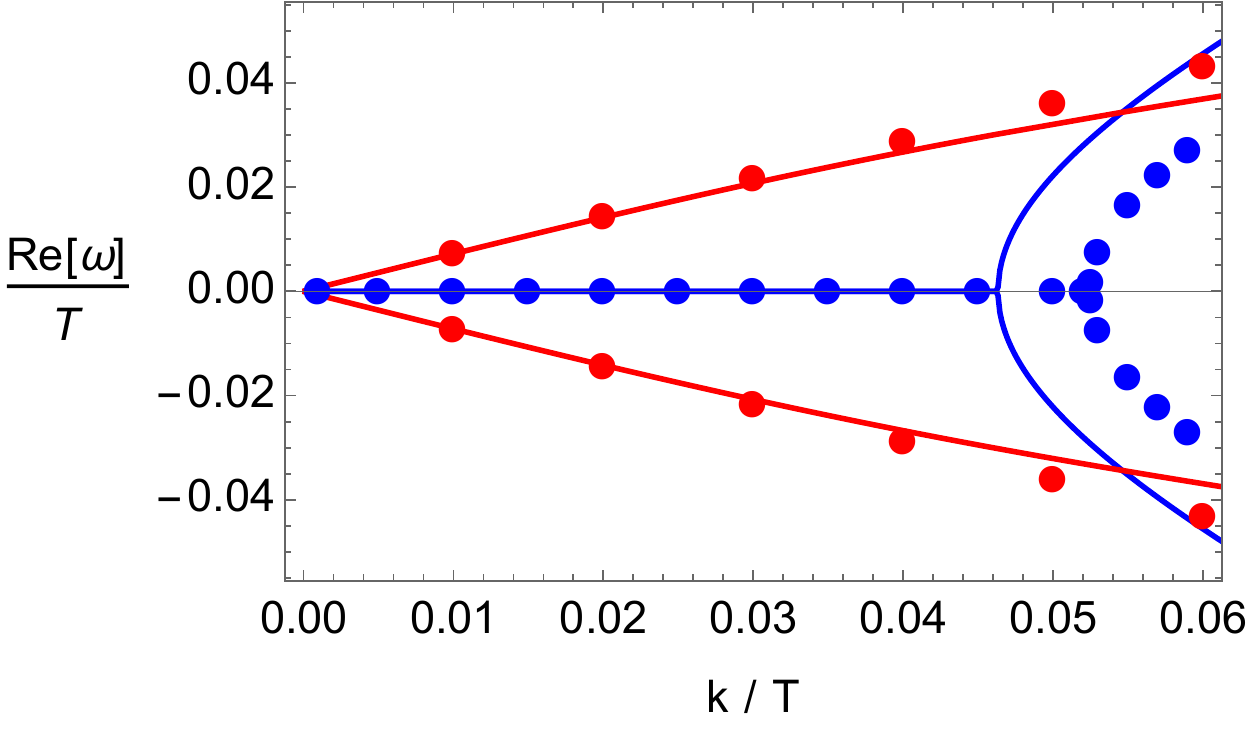} \label{}}\quad
     {\includegraphics[width=7cm]{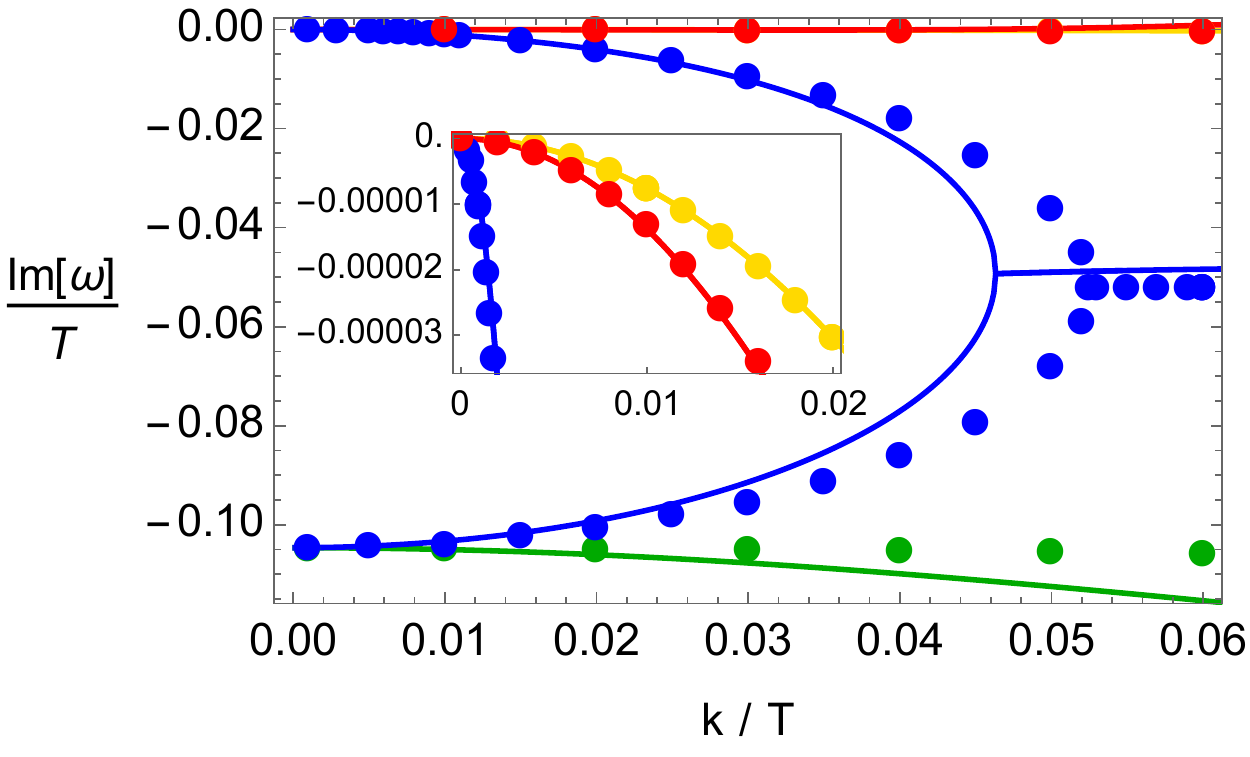} \label{zeroden3d}}
 \caption{Dispersion relations of the lowest QNMs at zero density ($\mu/T=0$) and $B/T^2 =1$. The symbols are the numerical results. The solid lines are the independent predictions from magnetohydrodynamics. Here, blue corresponds to the EM wave, red to the magnetosonic wave, yellow to shear diffusion, and green to the damped charge density fluctuations.}\label{fig2}
\end{figure}

In order to validate further our holographic model, we move to the description of the boundary field theory dynamics at zero chemical potential but finite magnetic field. In this case, the dynamics of the different fluctuations couple and the resulting low-energy modes are more interesting. 
The standard longitudinal sound, typical of a neutral relativistic fluid, is now modified into the so-called \textit{magnetosonic wave} whose dispersion is given by:
\begin{align}\label{SWSHW2}
\omega = \pm v_{\text{ms}} \, k - i \frac{\Gamma_\text{ms}}{2} \, k^2\,,
\end{align} 
where $v_{\text{ms}},\Gamma_\text{ms}$ are $B$-dependent parameters whose concrete expression can be found in \cite{Ahn:2022azl}. At the same time, the shear diffusion mode is modified in the limit of small magnetic field, $B/T^2 \ll 1$, into
\begin{align}\label{SWSHW2}
\begin{split}
\omega = -i \left( \frac{\eta}{\epsilon+p} - \frac{\eta B^2}{\mu_{\text{m}}(\epsilon+p)^2} \right) \, k^2 \,.
\end{split}
\end{align} 

\begin{figure}[]
\centering
     {\includegraphics[width=7cm]{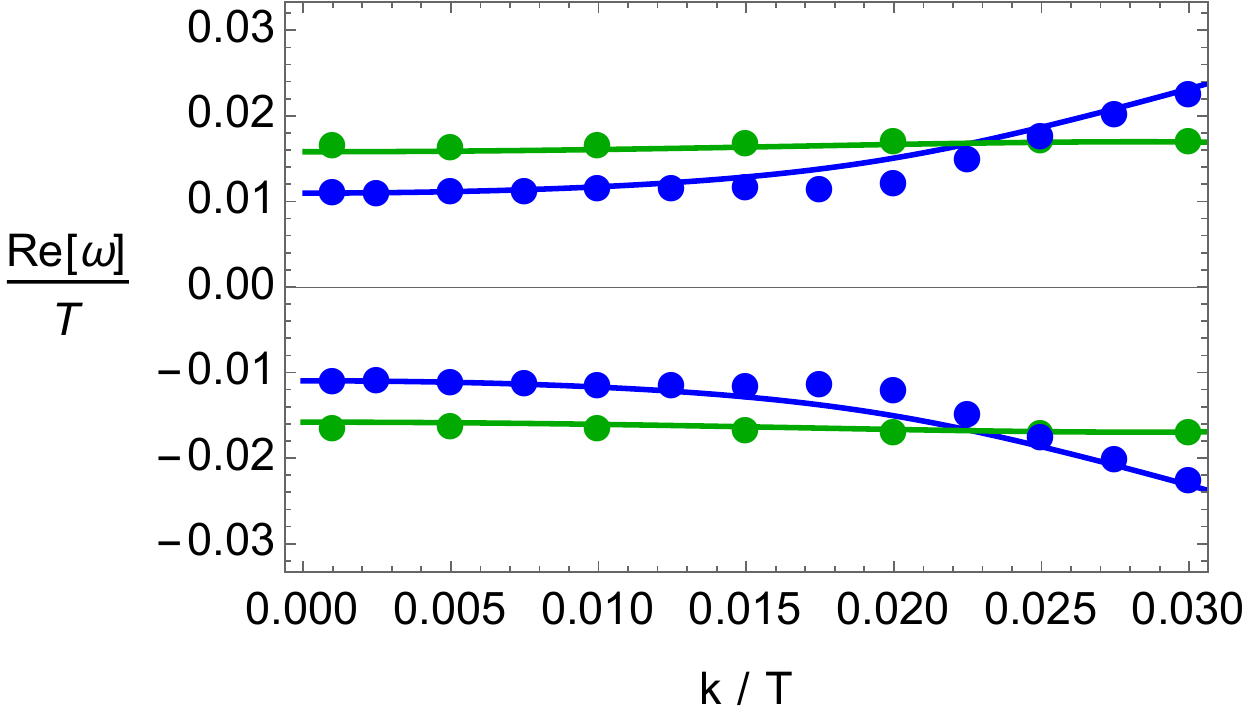} \label{finden1c}}\quad 
     {\includegraphics[width=7cm]{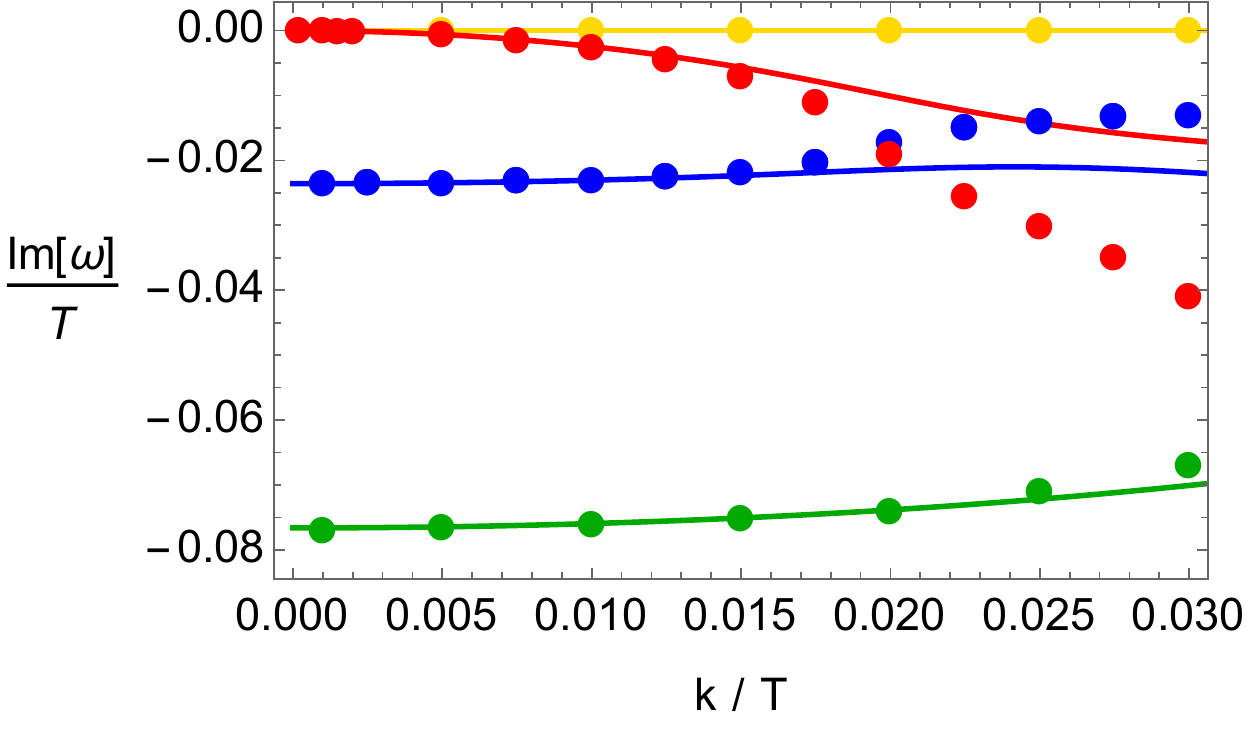} \label{finden1d}}
 \caption{Dispersion relations of the lowest QNMs at finite density ($\mu/T=0.5$) and $B/T^2=0.5$. The symbols are the numerical results. The solid lines are the independent predictions from magnetohydrodynamics. Red and yellow are the two hydrodynamic modes in Eq.\eqref{FINITEDEN1}. Green and blue are the non-hydrodynamic modes from Eq.\eqref{GAPEOMD}.}\label{fig3}
\end{figure}

The dynamics of the EM wave is qualitatively similar to the $B=0$ case, where now the parameters become as well function of $B$,
\begin{align}\label{EMWAVE2}
\omega \left(\omega + i \, \Sigma(B)\right) = v^2(B)\,k^2 \,.
\end{align} 
The expressions for $\Sigma(B),v(B)$ can be derived analytically within the magnetohydrodynamic framework and can be found in \cite{Ahn:2022azl}.
Finally, charge fluctuations do not diffuse anymore but rather relax as
\begin{align}\label{DCDD2}
\begin{split}
\omega = -i \, \left(\frac{\sigma}{\epsilon_{\text{e}}} + \frac{\sigma B^2}{\epsilon_{\text{e}} \, \mu_{\text{m}} (\epsilon+p)} \right) - i \, \left(\frac{\sigma}{\chi_{\rho\rho}} + \frac{\eta B^2}{\mu_{\text{m}}(\epsilon+p)^2} \right)\, k^2 \,,
\end{split}
\end{align} 
where $\chi_{\rho\rho} \equiv \partial \rho / \partial \mu$ is the charge susceptibility, and the limit of small magnetic field is assumed. In Fig.\ref{fig2}, we demonstrate that the numerical results obtained from the holographic model confirm the validity of magnetohydrodynamics even at finite magnetic field.

We can push this even further and switch on both a chemical potential and a magnetic field. In this regime, we are left with two hydrodynamic modes and four non-hydrodynamic modes. We find one longitudinal diffusive mode and one transverse subdiffusive shear mode with dispersions
\begin{align}\label{FINITEDEN1}
\omega =    -i \, \frac{ \, \rhomu (\epsilon+p)^2 \,\, \sigma}{T \, \left[ \eT\rhomu - \emu\rhoT \right] \left(\rho^2+B^2 \sigma\right)} \,\, k^2 \,, \qquad  \omega = -i \frac{\eta}{\mu_{\text{m}} \, \rho^2} \, k^4  \,.
\end{align} 

The remaining gapped four modes are obtained by solving the following equation:
\begin{align}\label{GAPEOMD}
\begin{split}
\left[ \omega \left( \omega + i \, \frac{\sigma}{\epsilon_{\text{e}}} \right) - \Omega_p^2 \right]^2 \,=\, \frac{B^2}{\epsilon_{\text{e}}^2 \, \mu_{\text{m}}^2(\epsilon+p)^2} \left[ \rho^2 - \mu_{\text{m}}^2 \sigma^2 (\rho^2-B^2)  +\omega^2\left( 2(\epsilon+p)(\sigma - i \epsilon_{\text{e}} \omega) \right)  \right]  \,,
\end{split}
\end{align} 
where $\Omega_p$ is the plasma frequency
\begin{align}\label{PSFOUR}
\begin{split}
\Omega_p^2 \,:=\, \frac{\rho^2}{\epsilon_{\text{e}}(\epsilon+p)}\,.
\end{split}
\end{align}

In Fig.\ref{fig3}, we show the agreement between the magnetohydrodynamic expressions and the numerical data obtained from the holographic QNMs. Within the (quasi-)hydrodynamic limit, roughly given by the condition $\omega(k=0)/T \ll 1$, the agreement is excellent.\\

In conclusion, 
\begin{enumerate}
    \item By carefully studying the dispersion relation of the low-energy modes, we have demonstrated that mixed boundary conditions allow for the existence of a boundary dynamical gauge field obeying Maxwell's equations. In colloquial terms, Sir James Clerk Maxwell is now sitting on the boundary of our holographic model.
    \item We have proved that the low-energy description of the boundary field theory is in excellent agreement with magnetohydrodynamics, confirming that the dual system is a strongly coupled plasma with dynamical electric and magnetic field and finite Coulomb interactions.
    \item We have shown that higher-form bulk fields are not necessary to discuss magnetohydrodynamics in the context of bottom-up holography. This is not surprising given the results of \cite{DeWolfe:2020uzb}.
\end{enumerate}

\section{A bona-fide holographic superconductor}
\subsection{Setup}
We consider the Einstein-Abelian-Higgs model in four spacetime dimensions,
\begin{equation}\label{GENMODELXX2222}
\begin{split}
S_{\text{bulk}} = \int d^4x \sqrt{-g} \left[ R + 6 - \frac{1}{4} F^2 -|D\Psi|^2 \, -M^2 |\Psi|^2  \right]  \,,
\end{split}
\end{equation}
where we have defined the covariant derivative $D_{\mu} := \partial_{\mu} -i q A_{\mu}$, the charge $q$ of the complex bulk scalar field $\Psi$ and its mass $M$. The mass and the charge will be respectively fixed to $q=1$ and $M^2=-2$. For simplicity, we work in the probe limit and only consider a fixed background metric given by:
\begin{equation}\label{METABG}
d s^2 =  \frac{1}{z^2}\left(-f(z)\, d t^2 +  \frac{dz^2}{f(z)}  +  d x^2 +  d y^2 \right) \,,\quad f(z) = 1 - \frac{z^3}{z_{h}^3}\,,
\end{equation}
with $z_h$ the location of the black hole horizon and $z=0$ the UV boundary of the spacetime.
The corresponding temperature and entropy density of the dual field theory are given by:
\begin{equation}
    T = \frac{3}{4 \pi z_{h}}\,,\quad s= \frac{4\pi}{z_h^{2}} \,.
\end{equation}
We then assume the following Ansatz for the matter fields
\begin{equation}\label{PROBEFIELDSEQ}
A =  A_{t}(z) \, d t \,,  \qquad \Psi = \psi(z)\,.
\end{equation}
which obey the following asymptotic behaviors 
\begin{equation}\label{BGADSEX}
\begin{split}
A_{t} = \mu - \rho z + \mathcal{O} (z^2) \,, \quad \psi = \psi_1 z + \psi_2 z^2 + \mathcal{O} (z^3) \,
\end{split}
\end{equation}
near the AdS boundary located at $z=0$. Here, $\mu,\rho$ are respectively the chemical potential and the charge density of the dual field theory. Additionally, using Dirichlet boundary conditions for the charged bulk scalar field (standard quantization), $\psi_1$ is interpreted as the source for a charged scalar operator with conformal dimension $\Delta=2$, and $\psi_2$ gives its vacuum expectation value, the condensate $\langle 
\mathcal{O}_2\rangle$. In the following, we will be interested in solutions which break spontaneously the U(1) symmetry, and therefore correspond to the choice $\psi_1=0$ and $\psi_2\neq 0$. These solutions can be easily constructed numerically. The broken phase appears below a critical temperature which we denote as $T_c$.

On top of our background, we consider the following fluctuations:
\begin{equation}\label{}
\begin{split}
    \delta A &=  \delta a_{t}(t, z, x) \, d t + \delta a_{x}(t, z, x) \, d x + \delta a_{y}(t, z, x) \, d y \,, \\ 
\delta \Psi &=  \delta \sigma(t, z, x) + i \, \delta \eta(t, z, x) \,,
\end{split}
\end{equation}
where the radial gauge $A_{r}=0$ is assumed. Importantly, while for the scalar fluctuations we retain the classical Dirichlet boundary conditions, the boundary conditions for the gauge field fluctuations will be modified as in the previous section in order to make the boundary gauge field dynamical. For details, we refer to \cite{Jeong:2023las}.

We now jump to the study of the low-energy collective excitations in the dual field theory to demonstrate the existence of a bona-fide superconducting state. Before proceeding, let us just quickly remind the reader what happens in the case of a superfluid with non-dynamical (global) U(1) symmetry. In that case, the superfluid phase displays two pairs of sound modes (only one if the probe limit is taken) and a damped amplitude mode. One of the sound modes, known as second sound, is an immediate consequence of the existence of a superfluid Goldstone mode, and its velocity is proportional to the superfluid density. In the context of holographic superfluids, the low-energy dynamics has been matched 1-to-1 with relativistic superfluid hydrodynamics in \cite{Arean:2021tks}. Here, we initiate a similar analysis for the holographic superconductor, starting from the probe limit.

\begin{figure}[]
\centering
   \includegraphics[width=6cm]{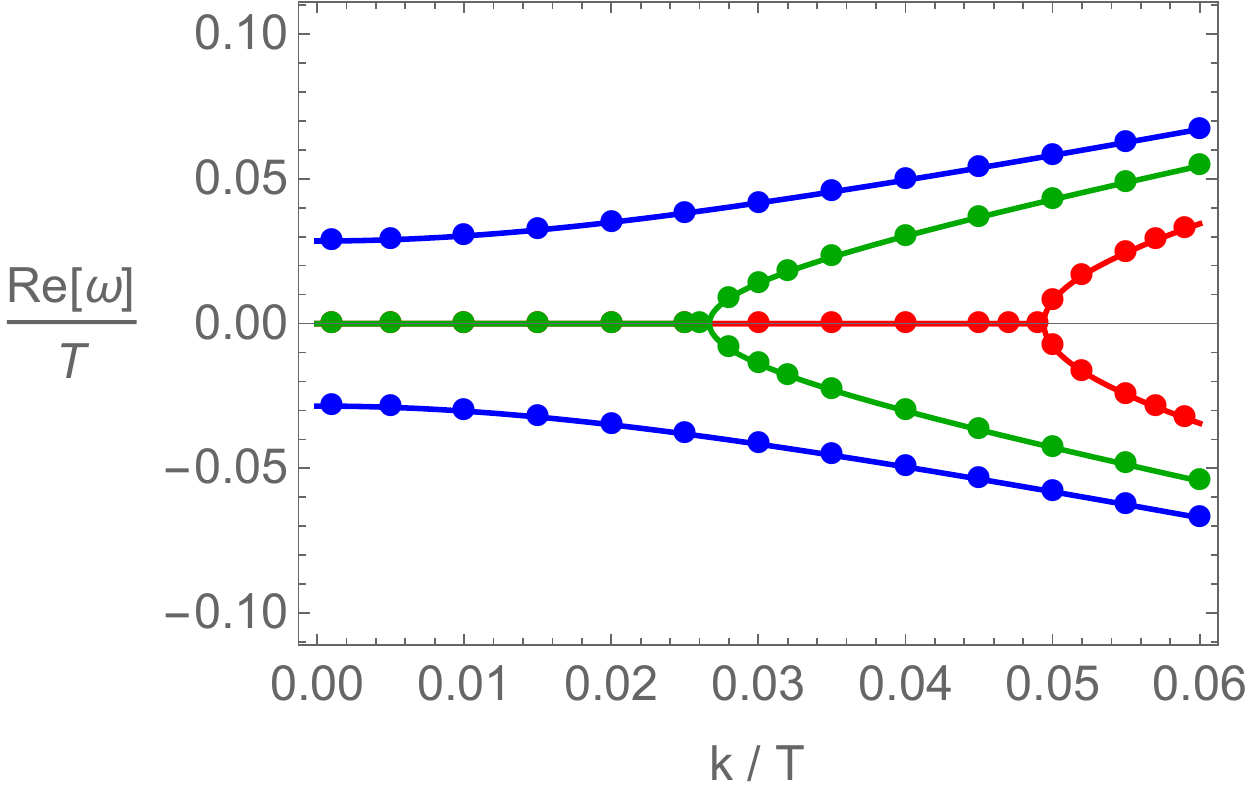} \quad
     \includegraphics[width=6cm]{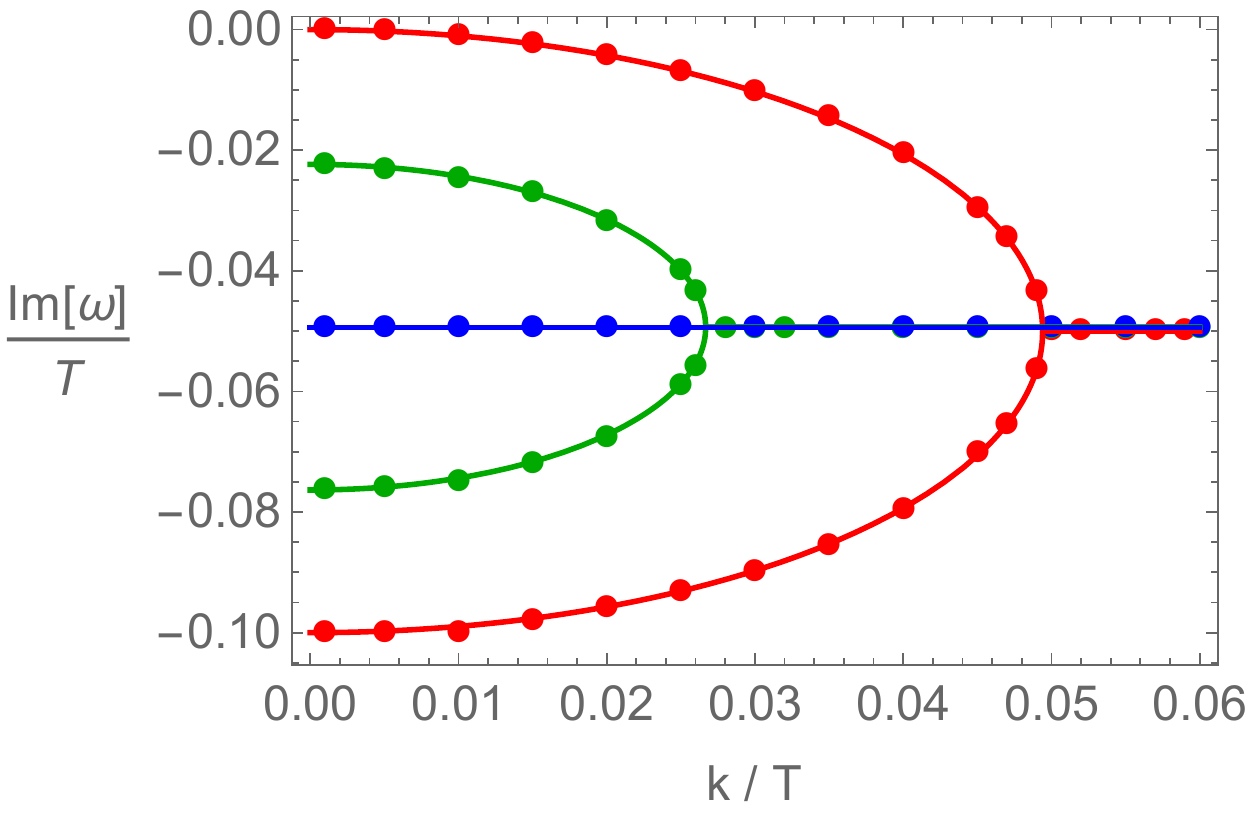} 
     \vspace{0.2cm}
     
     \includegraphics[width=6cm]{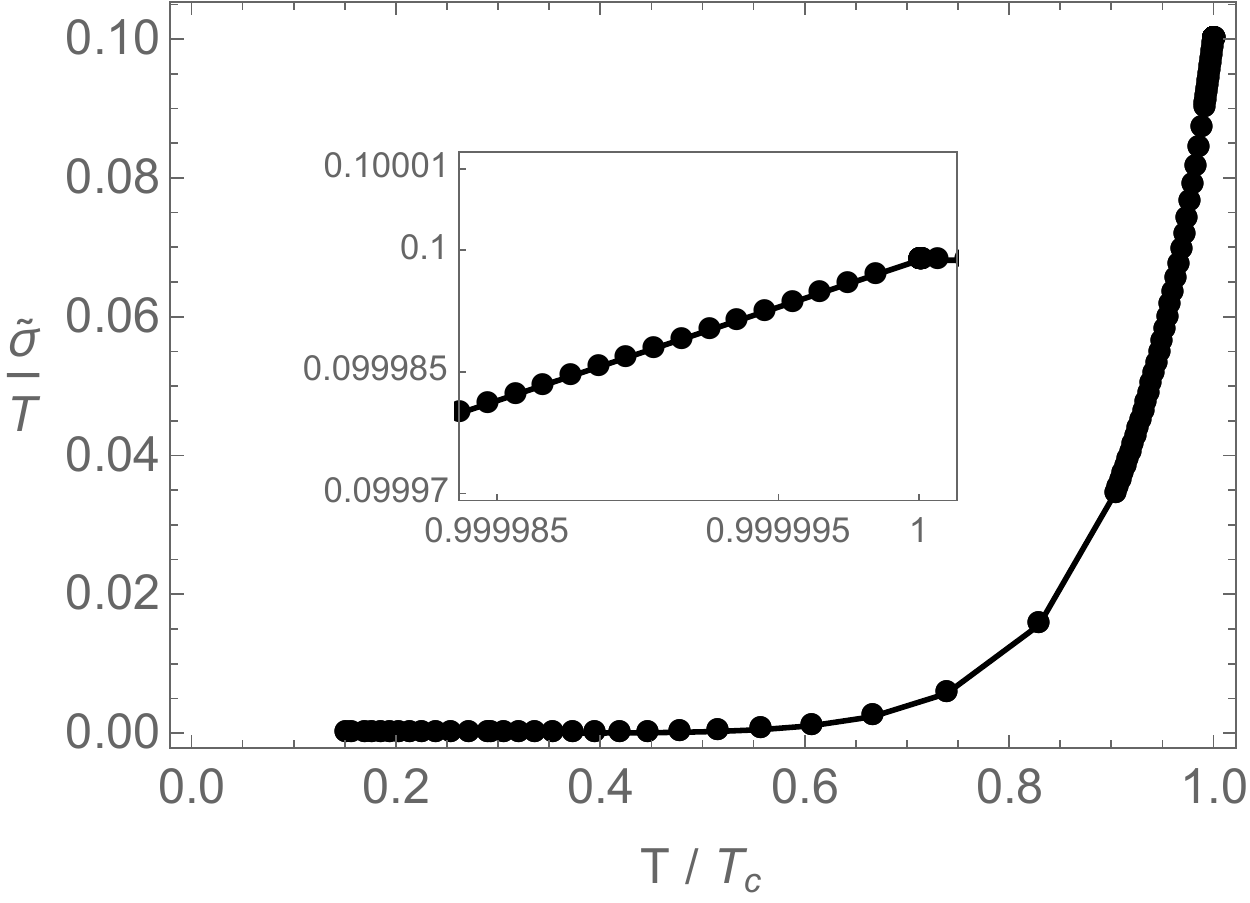} 
     \quad
      \includegraphics[width=6cm]{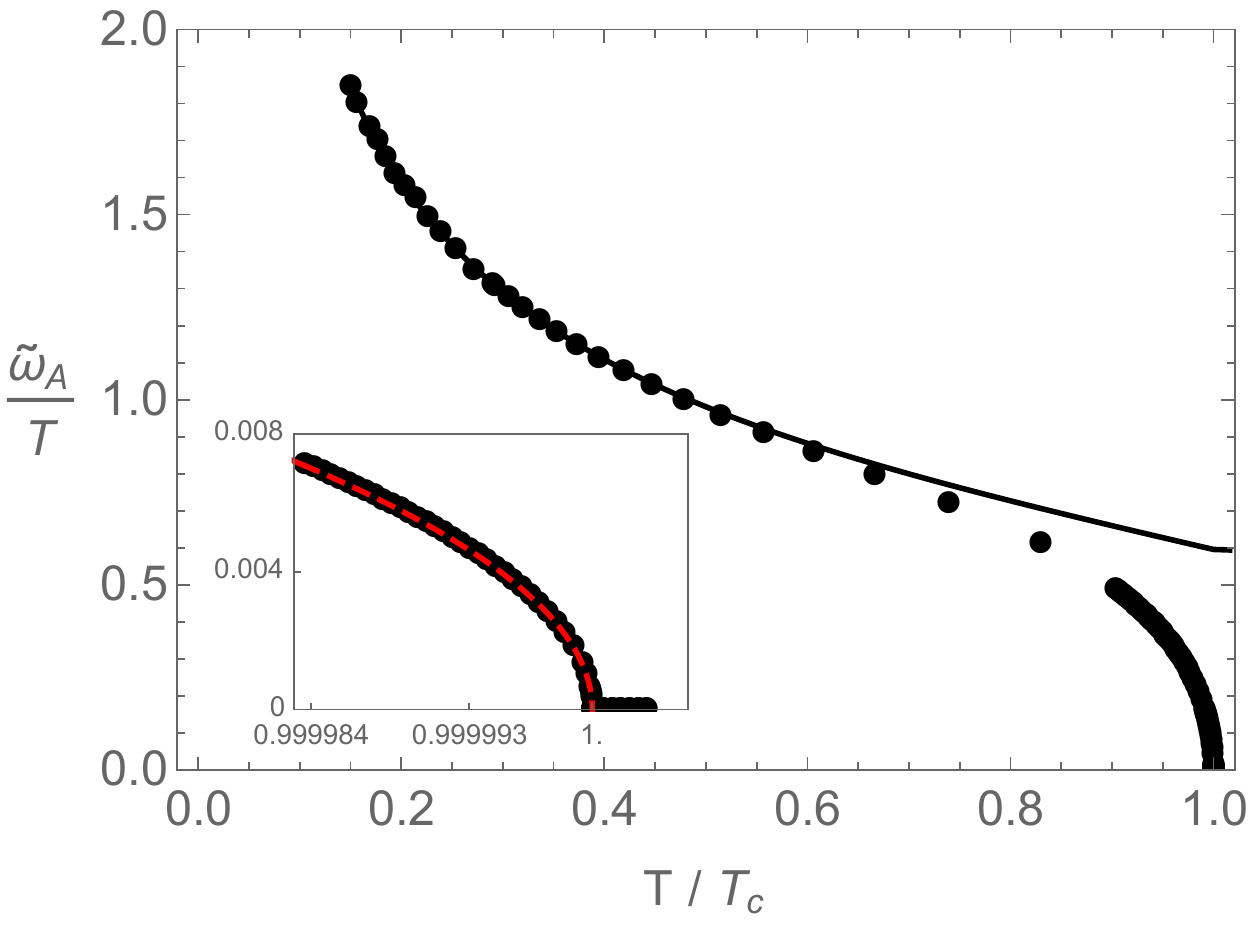} 
 \caption{\textbf{(Top panel)} The dispersion relation of the lowest collective modes in the transverse sector for different values of the reduced temperature $T/T_c = (1, 0.999, 0.998)$ (red, green, blue). Symbols represent the numerical values and solid lines are fits using Eq.\eqref{TRANSFIT1}. \textbf{(Bottom panel)} The temperature dependence of the phenomenological parameters $\tilde \sigma$ and $\tilde \omega_A$. Dots are evaluated from the numerical fits. Solid lines represent the analytical expression in Eq. \eqref{TRANSFIT2} (left panel) and the plasma frequency value (right panel). The insets show the data near the critical point, $T=T_c$, Eq.\eqref{TRANSFIT4}.}\label{fig4}
\end{figure}
\subsection{Highlights}
In this section, we present the most important results of our analysis. For a complete and in detail description, see \cite{Jeong:2023las}. Let us start from the transverse sector of fluctuations. In order to understand the dynamics in the transverse sector, and taking advantage of the intuition coming from Ginzburg-Landau theory, we utilise the following equation: 
\begin{align}\label{TRANSFIT1}
\omega^2 = \tilde{\omega}_A^2 + \tilde{v}^2 k^2 - i \, \tilde{\sigma} \, \omega \,.
\end{align}
In the normal phase, $T>T_c$, the dispersion of the low energy modes is predicted by the magneto-hydrodynamic theory discussed in the previous section and studied in detail in \cite{Ahn:2022azl}. In particular, at zero charge density and zero magnetic field, we expect the transverse fluctuations to be governed by equation \eqref{EMWAVE}. The latter corresponds to the limit $\tilde{\omega}_A=0$ in Eq.\eqref{TRANSFIT1}.\\
In the broken (superconducting) phase, the transverse excitations develop a ``mass'' term $\tilde{\omega}_A$. Close to the critical point, such a mass term obeys a scaling relation 
\begin{align}\label{TRANSFIT4}
\begin{split}
\tilde{\omega}_A \,=\, \alpha \, \sqrt{1-T/T_c} \,,
\end{split}
\end{align}
which can be derived using the Ginzburg-Landau formalism. Interestingly, one can also obtain a perturbative analytical result in terms of bulk quantities, given by
\begin{align}\label{MAKOTOFOR}
\tilde{\omega}_A \,=\, \sqrt{\frac{2\lambda}{1+\lambda} \, I} \,, \qquad I := \int_{0}^{1} d z \, \left(\frac{\psi(z)}{z}\right)^2 \,,
\end{align}
and derived in Ref.\cite{Natsuume:2022kic}, where for simplicity we have fixed $z_h=1$. In \cite{Jeong:2023las}, we proved that such a formula is in excellent agreement with the numerical results.

On the contrary, at low temperature, approximately below $T/T_c\approx 0.5$, the frequency $\tilde{\omega}_A$ coincides with the plasma frequency. These results are shown explicitly in Fig.\ref{fig4}. The top panel shows the agreement between the numerical modes at finite wave-vector and the solutions of Eq.\eqref{TRANSFIT1}. At low temperature (blue data points), the modes exhibit the expected massive dispersion. In the bottom left panel of Fig.\ref{fig4}, we show that the parameter $\tilde{\sigma}$ is identified with the conductivity:
\begin{align}\label{TRANSFIT2}
\tilde{\sigma} \,=\, \sigma_0 \, \lambda \,,\qquad \sigma_0 := \lim\limits_{\omega \to 0} \text{Re}[\sigma(\omega)]\,.
\end{align}
which is extracted numerically using the standard Kubo formula and holographic methods (see \cite{Baggioli:2019rrs} for a pedagogical explanation and an open-source code).

\begin{figure}[]
\centering     
     \includegraphics[width=6cm]{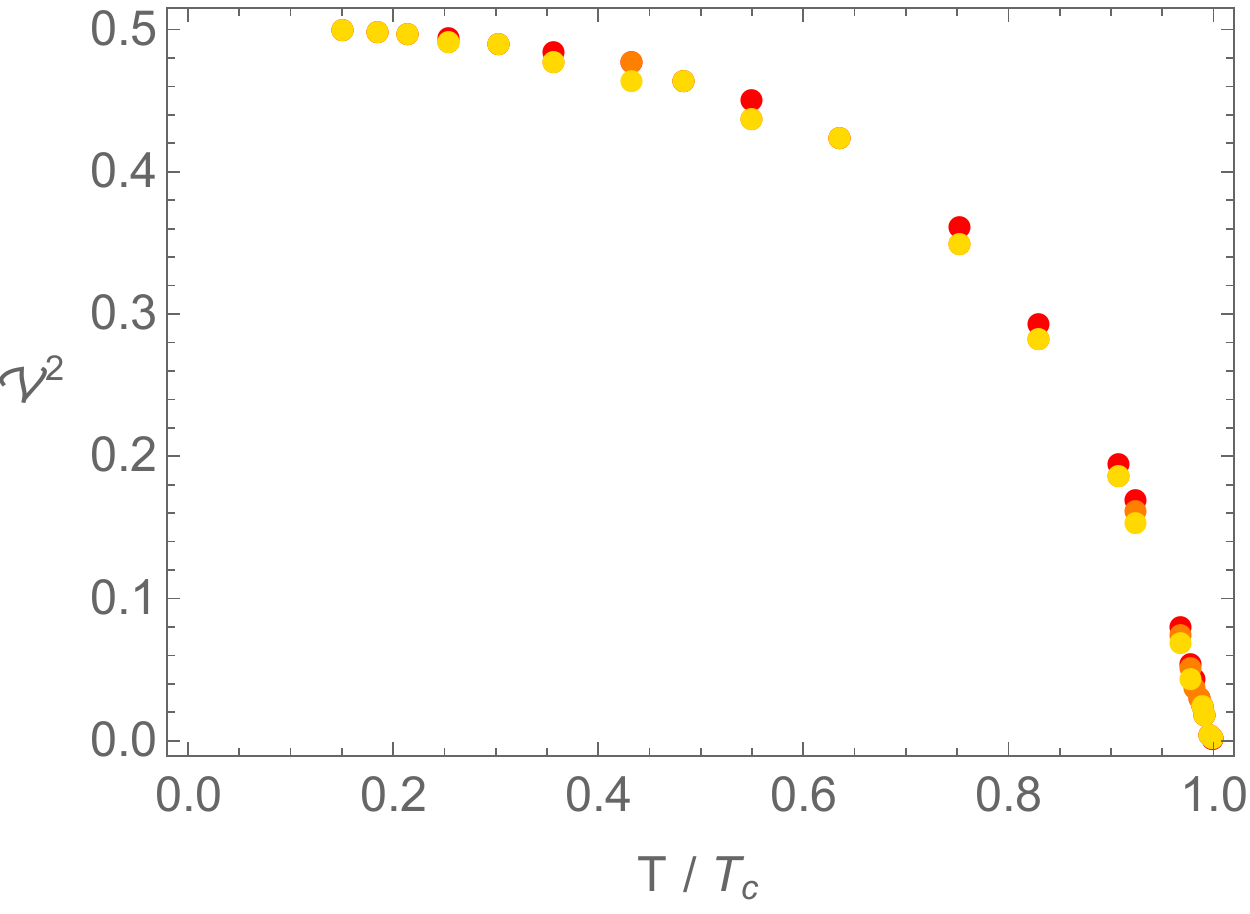} \quad
     \includegraphics[width=6cm]{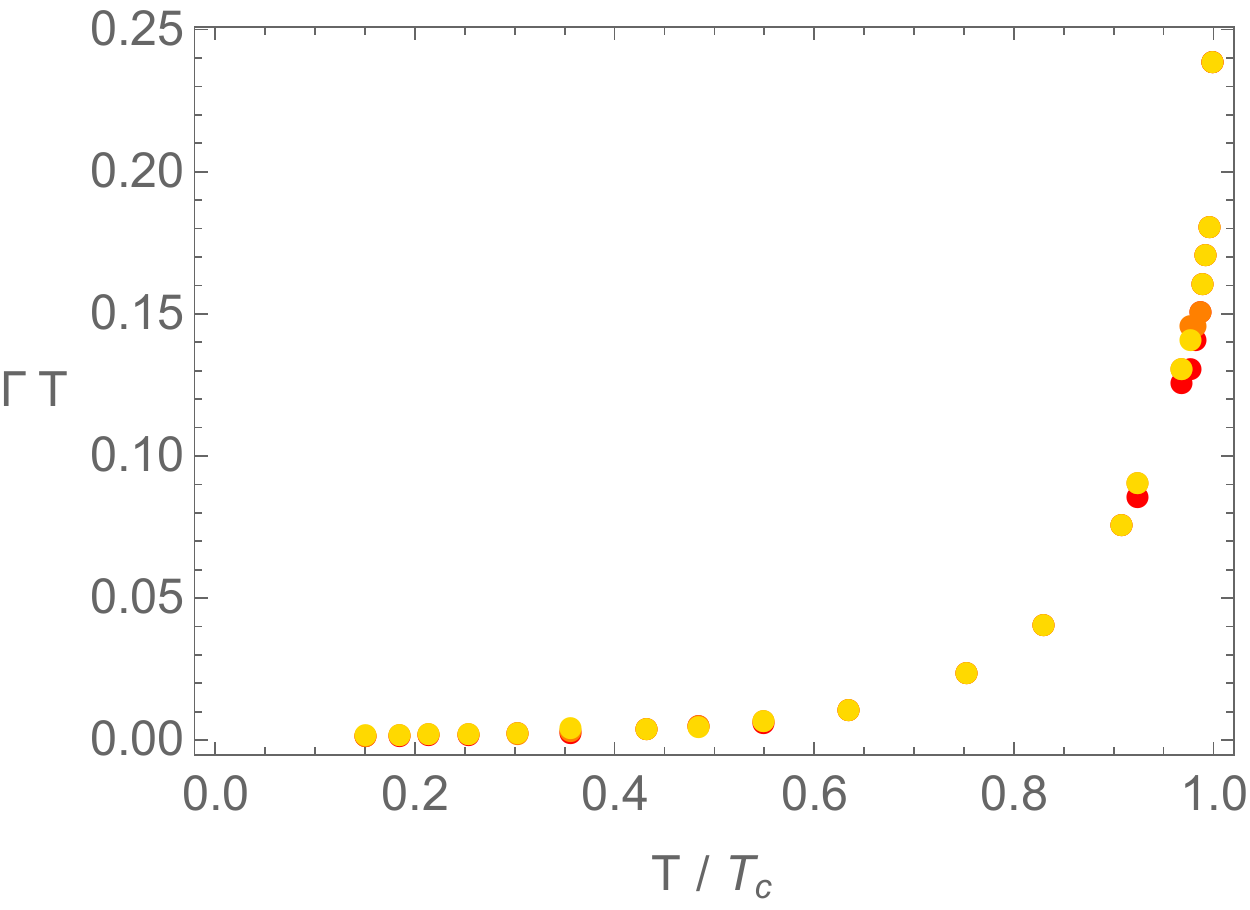} 
     
 \caption{The parameters $\mathcal{V},\Gamma$ as a function of the reduced temperature for $\lambda = 0.1, 0.2, 0.3$ (red, orange, yellow).}\label{PLOTFW4}
\end{figure}

Moving to the longitudinal sector, the dynamics becomes quite complex. A complete effective description (either using quasi-hydrodynamics or some Ginzburg-Landau like formalism) has not been obtained yet. Nevertheless, at least in the low wavevector limit, the dispersion relations of the low energy modes are well described by the following decoupled equations
\begin{align}\label{LONGFOR1}
\begin{split}
\omega  \left(\omega + i \, \tilde{\sigma} + i \, \Gamma \, k^2 \right) = \mathcal{V}^2 \, k^2 + \tilde{\omega}_A^2 \,, \qquad \omega + i \, \Omega + i \, D_{\Omega} \, k^2 = 0 \,.
\end{split}
\end{align}
Importantly, the only dependence on the electromagnetic interactions, $\sim \lambda$, appears in the parameters $\tilde \sigma, \tilde \omega_A$, which are the same as the ones discussed in the transverse sector. In absence of dynamical EM, both of them vanish and we recover the standard low-energy modes for a relativistic superfluid. In particular, the left equation in \eqref{LONGFOR1} would give rise to the superfluid second sound, with velocity $\mathcal{V}$ and attenuation constant $\Gamma$. On the contrary, the right equation in \eqref{LONGFOR1} gives the dynamics of the amplitude/Higgs mode. Notice that, for finite EM interactions (i.e., in a superconductor), the hydrodynamic second sound mode disappears from the spectrum as it is ``eaten'' by the dynamical gauge field. Indeed, from Eq.\eqref{LONGFOR1} no hydrodynamic mode (whose frequency goes to zero as $k\rightarrow 0$) is expected. This phenomenon is nothing else than the famous Higgs mechanism, which is now naturally realized in the holographic model thanks to the modified boundary conditions.

\begin{figure}[]
\centering
     \includegraphics[width=6cm]{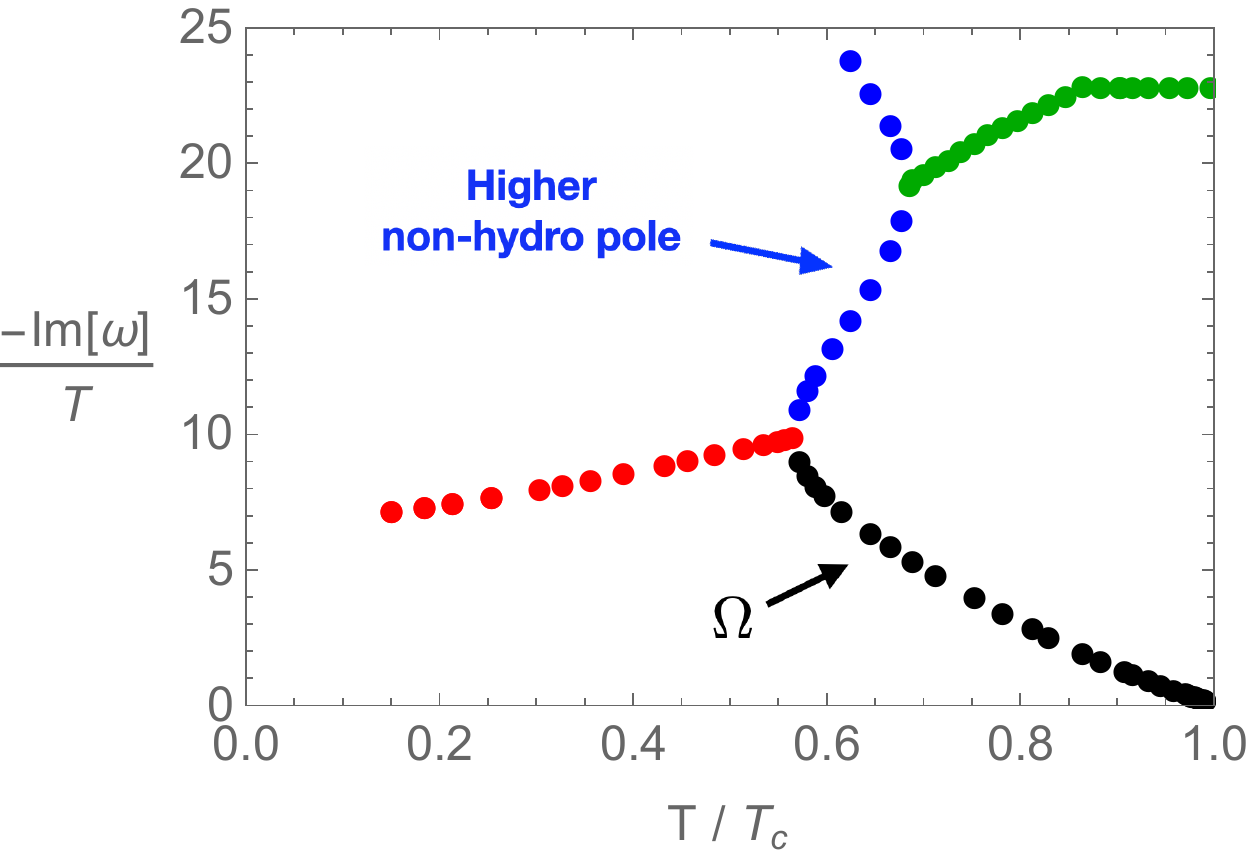} 
     \includegraphics[width=6.0cm]{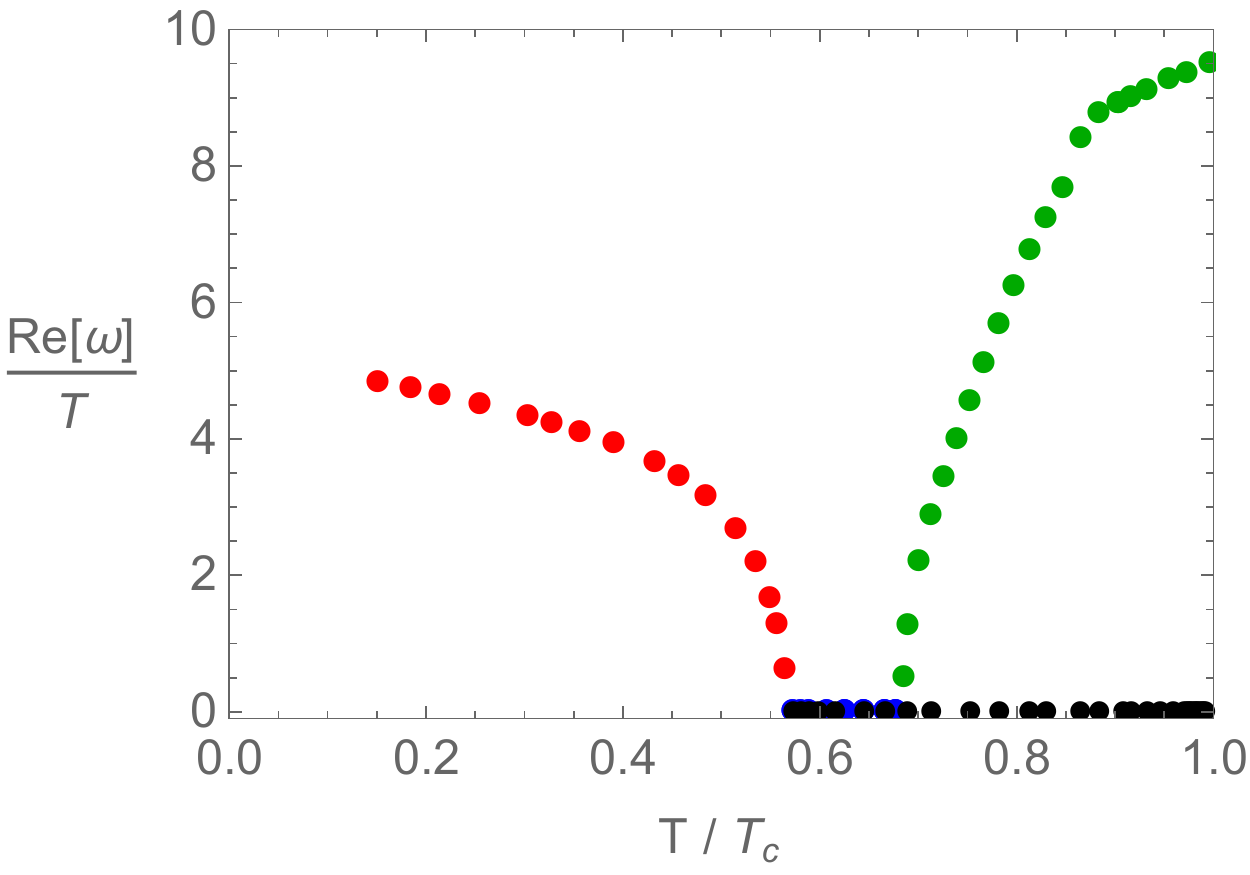} 

     \vspace{0.2cm}
     
      \includegraphics[width=6.0cm]{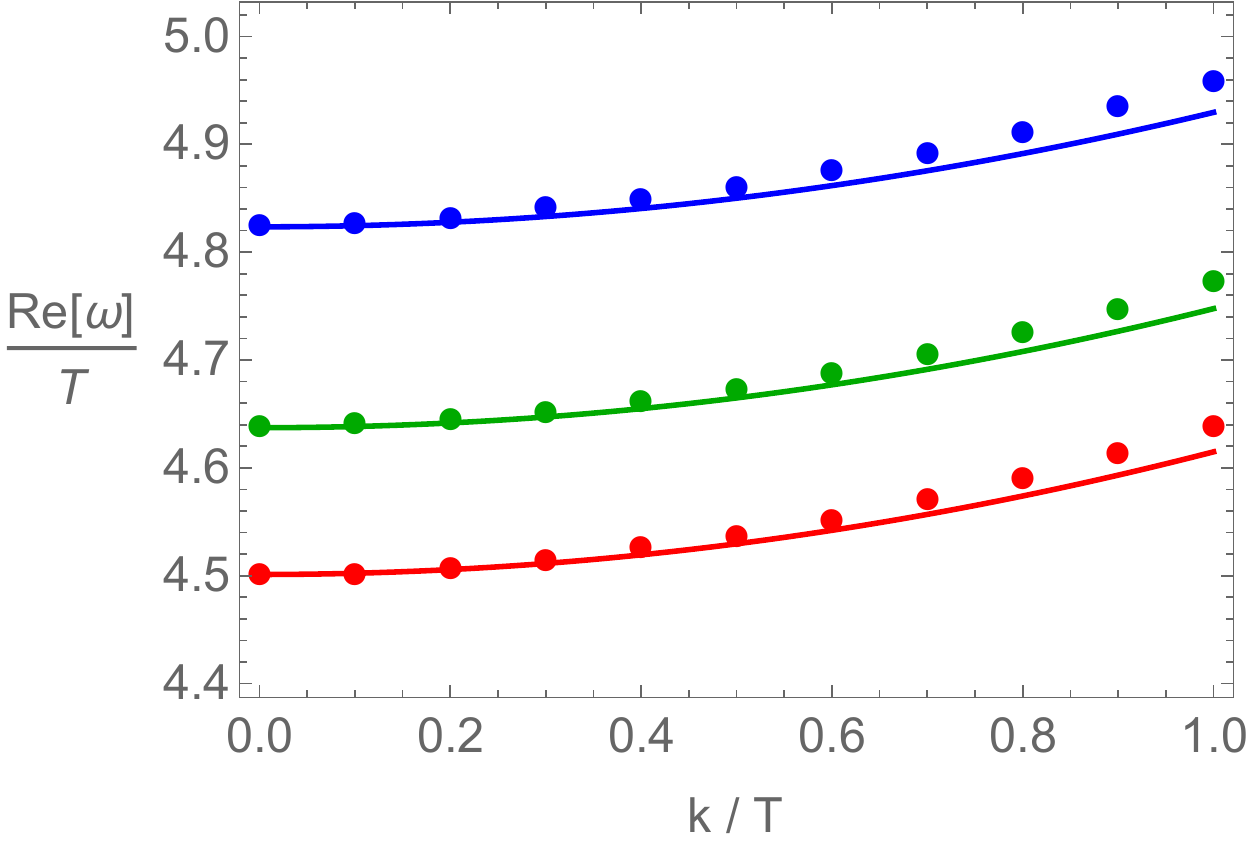} \quad
     \includegraphics[width=6cm]{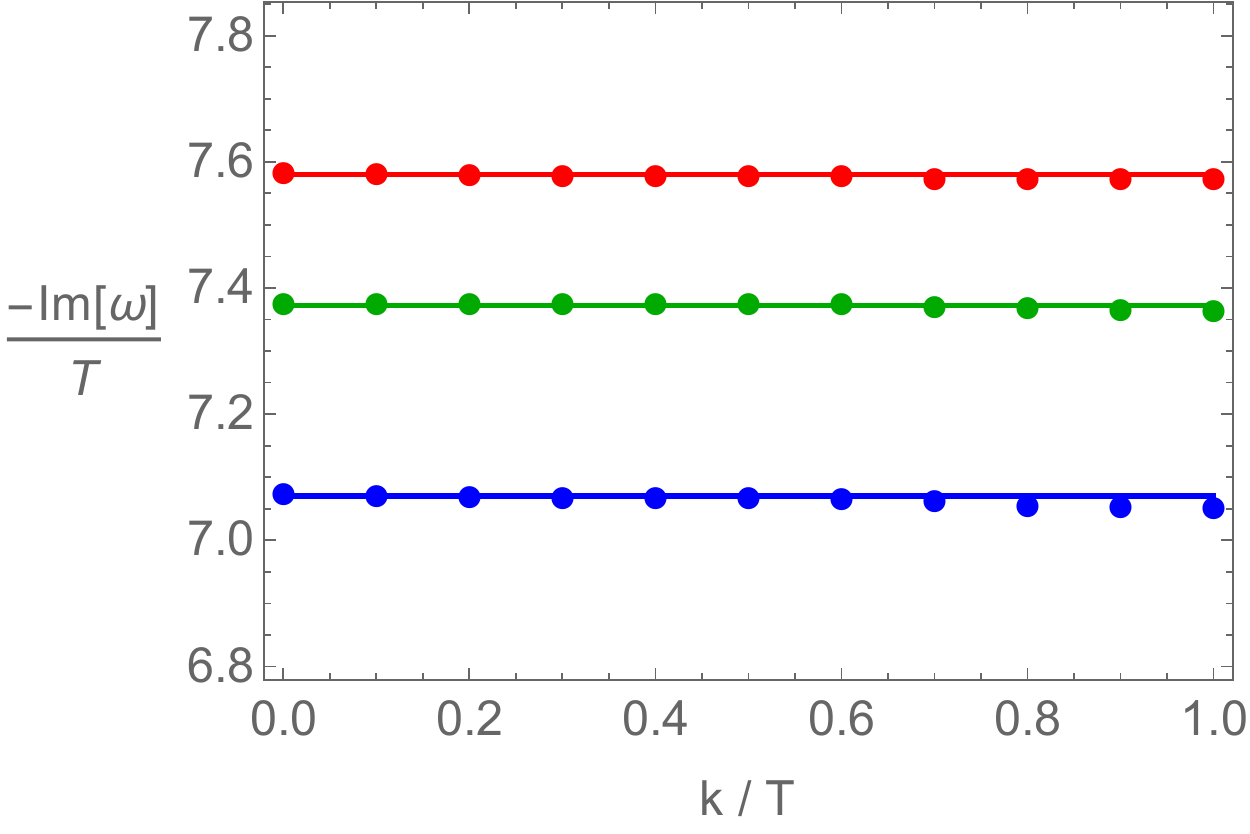} 
 \caption{\textbf{(Top panel)} The dynamics of the Higgs mode as a function of the reduced temperature. \textbf{(Bottom panel)} The dispersion relation of the Higgs mode at low temperature: $T/T_c = (0.15, 0.21, 0.25)$ (blue, green, red).}\label{PLOTFW2}
\end{figure}

In \cite{Jeong:2023las}, we found that Eq.\eqref{LONGFOR1} provides a good description of the low energy dynamics in the holographic model in perfect agreement with the numerical data. Interestingly, we find that the parameters $\Gamma,\mathcal{V},\Omega,D_\Omega$ are independent of the EM coupling $\lambda$ and their value coincides with that reported in the holographic superfluid model. As a demonstration, the behavior of $\Gamma,\mathcal{V}$ is shown in Fig.\ref{PLOTFW4} for different values of the EM coupling. Finally, as expected from GL theory, and demonstrated analyiticslly using holography \cite{Donos:2022xfd}, we find numerically that:
\begin{equation}
    \Omega\sim(1-T/T_c)\,.
\end{equation}
Decreasing the temperature, we find that such overdamped mode (black dots in the top panel of Fig.\ref{PLOTFW2}) collides with a second non-hydrodynamic mode on the imaginary axes and creates a pair of complex modes with increasing real part (red dots in the top panel of Fig.\ref{PLOTFW2}). This dynamics is well predicted by Ginzburg-Landau theory (see \cite{Jeong:2023las} for details). We find that the crossover between overdamped and underdamped dynamics appears approximately at $T/T_c \approx 0.6$, and its location is importantly independent of the EM coupling $\lambda$. 

At low temperature, we clearly observe a massive mode with an attenuation constant approximately indendepent of the wave-vector (see bottom panel of Fig.\ref{PLOTFW2}), which we interpret as the emergent massive Higgs mode expected in a superconducting system. Interestingly, a similar mode was observed in the holographic superfluid model \cite{Bhaseen:2012gg}, but the corresponding dynamics was profoundly different. As a matter of fact, the massive mode appears therein as a highly non-hydrodynamic mode (not related to the fluctuations of the amplitude) which becomes long-lived in the limit of small temperature.

Finally, we investigate the size of the mass of the Higgs mode at small temperature. We find numerically that, 
\begin{equation}\label{ele}
 \frac{\tilde{\omega}_{H}}{2\Delta}\Big|_{T\approx 0.15 T_c} \approx 0.162  \,,
\end{equation}
where $\Delta$ is the superconducting energy gap related to the order parameter as $2\Delta = \sqrt{\langle O_{2} \rangle}$ \cite{Hartnoll:2008vx}. $T\approx 0.15 T_c$ is the lowest temperature value that we can trust within the probe limit. The value in Eq.\eqref{ele} is an order of magnitude smaller than the usually reported one in BCS-type superconductors \cite{doi:10.1146/annurev-conmatphys-031214-014350,Shimano:2019iqf}. \\

In conclusion,
\begin{enumerate}
    \item We have studied the low-energy collective modes in a bona-fide holographic superconductor and showed the existence of the Higgs mechanism. In colloquial terms, Sir Peter Higgs is sitting on the boundary of our holographic model.
    \item We have highlighted, within the holographic framework, the differences between the holographic superfluid model and the bona-fide holographic superconductor. In particular, we have demonstrated that the gapless second sound mode disappears from the spectrum and that a massive Higgs mode emerges at low temperature.
\end{enumerate}

\section{Notes to the future}
We have shown the power of modified boundary conditions in order to introduce dynamical EM interactions into the dual boundary field theory of simple bottom-up holographic models. We have verified the validity of this procedure by considering two concrete applications: (I) a $2+1$-dimensional strongly coupled plasma described by relativistic magnetohydrodynamics, and (II) a $2+1$ superconductor exhibiting Higgs mechanism.

Let us notice that a similar application of the same boundary conditions in the context of condensed matter systems regards the study of holographic plasmons, which has attracted a great deal of interest in the recent couple of years \cite{Gran:2017jht,Gran:2018iie,Gran:2018vdn,Gran:2018jnt,Baggioli:2019aqf,Baggioli:2019sio,Baggioli:2021ujk,Mauri:2018pzq,Romero-Bermudez:2018etn}. Indeed, some of the magnetohydrodynamic properties of the dual field theory, including the dispersion relations of the low-energy modes, have been already partially discussed therein.

We also notice that the usage of mixed boundary conditions is not restricted to the case of bulk gauge fields, but can also be promoted for the bulk metric, rendering the boundary geometry dynamical \cite{Compere:2008us,Ecker:2021cvz,Ishibashi:2023luz,Ecker:2023uea}.\\

Let us conclude this short manuscript with a few thoughts and questions for the future.
\begin{enumerate}
    \item The formulation of magnetohydrodynamics in the large magnetic field limit still presents open puzzles (see for example the recent work \cite{Vardhan:2022wxz}). The holographic model presented in \cite{Ahn:2022azl} is a perfect candidate to settle down the existing problems and ``help' the hydrodynamic and effective field theory sides to reach a correct and complete final framework.
    \item There is recent increasing interest on the formulation of chiral magnetohydrodynamics \cite{Landry:2022nog,Das:2022fho,Das:2022auy}. By combining the model in \cite{Ahn:2022azl} with the holographic setup for chiral fluids of \cite{Ammon:2020rvg}, one could (and should) obtain a holographic model for chiral magnetohydrodynamics.
    \item Inspired and helped by the holographic results of \cite{Jeong:2023las}, it would be interesting to construct a complete effective formalism to describe relativistic superconductors including dissipative effects. This would necessarily involve a generalization of hydrodynamics, in order to consider the gapped modes, and/or a generalization of the superfluid model F in the Hohenberg-Halperin classification \cite{RevModPhys.49.435} in presence of dynamical gauge fields. A holographic analysis on the lines of \cite{Donos:2022qao} would be certainly useful.
    \item Despite the equivalence between the higher-form language and the mixed boundary conditions has been clearly elucidated in \cite{DeWolfe:2020uzb}, when these theories are minimally coupled to matter fields the situation is less clear. It would be fruitful to understand better if the two can still be mapped into each other even in presence of matter. For example, how could one construct a bona-fide holographic superconductor using the higher-form language? The answer might be close to the results of \cite{Delacretaz:2019brr,Gursoy:2010kw,Hofman:2017vwr}.
\end{enumerate}
We are planning to work in these directions in the near future. 
\subsection*{Acknowledgments}
I would like to specially thank my friend and colleague Shingo Takeuchi for organizing this wonderful event. I would like to thank all the participants of the 6$^{th}$ International Conference on Holography, String Theory and Spacetime for fruitful and useful comments about our work and for interesting discussions about physics in general.
I thank my collaborators in \cite{Ahn:2022azl,Jeong:2023las} for fixing $2^n$ typos and mistakes in a preliminary version of this manuscript.
I acknowledge the support of the Shanghai Municipal Science and Technology Major Project (Grant No.2019SHZDZX01) and the sponsorship from the Yangyang Development Fund. I would like to thank Chuo University and RIKEN for the hospitality during the completion of this manuscript.

\bibliography{refs}

\end{document}